\documentclass[a4paper,fleqn,usenatbib,useAMS]{mnras}
\usepackage{graphicx}	
\usepackage{amsmath}	
\usepackage{amssymb}    
\usepackage{amssymb}	
\usepackage{multicol}	
\usepackage{bm}		
\usepackage{pdflscape}	
\usepackage{hyperref}

\usepackage{xcolor}

\newcommand*\mean[1]{\bar{#1}}

\newcommand{\rockstar}{{\tt ROCKSTAR}}
\newcommand{\sparta}{{\tt SPARTA}}

\newcommand{\hmsun}{h^{-1}\ {\rm M_{\odot}}}
\newcommand{\hMpc}{h^{-1}\ {\rm Mpc}}

\newcommand{\ximm}{\xi_{\rm mm}}
\newcommand{\xihm}{\xi_{\rm hm}}
\newcommand{\xihmone}{\xi_{\rm hm}^{\rm 1h}}
\newcommand{\xihmtwo}{\xi_{\rm hm}^{\rm 2h}}
\newcommand{\ximmtwo}{\xi_{\rm mm}^{\rm 2h}}
\newcommand{\xihh}{\xi_{\rm hh}}
\newcommand{\re}{r_{\rm e}}
\newcommand{\rt}{r_{\rm t}}
\DeclareMathOperator\erfc{erfc}
\newcommand{\reff}{r_{\rm eff}}
\newcommand{\Deltaeff}{\Delta_{\rm eff}}
\newcommand{\avg}[1]{\langle #1 \rangle}
\newcommand{\dsc}{\delta_{\rm sc}}
\newcommand{\rhos}{\rho_{\rm s}}
\newcommand{\rs}{r_{\rm s}}
\newcommand{\thetat}{\theta_{\rm t}}

\newcommand{\bPB}{b_{\rm PB}}

\title[Redefining the Halo Boundary]{A Redefinition of the Halo Boundary Leads to a Simple yet Accurate Halo Model of Large Scale Structure}
\author[Garcia et al.]{Rafael Garc\'ia$^1$\thanks{E-mail: rgarciamar@email.arizona.edu}, Eduardo Rozo$^{1}$, Matthew R. Becker$^{2}$, Surhud More$^{3,4}$
\\
$^{1}$Department of Physics, University of Arizona, Tucson, AZ 85721, USA \\
$^{2}$High Energy Physics Division, Argonne National Laboratory, Lemont, IL 60439, USA\\
$^{3}$The Inter-University Center for Astronomy and Astrophysics, Post bag 4, Ganeshkind, Pune, 411007, India\\
$^{4}$Kavli Institute for the Physics and Mathematics of the Universe (WPI), 5-1-5 Kashiwanoha, Chiba, 2778583, Japan\\
}  
\begin{document}

\maketitle 

\label{firstpage}

\begin{abstract}
We present a model for the halo--mass correlation function that explicitly incorporates halo exclusion.  We assume that halos trace mass in a way that can be described using 
a single scale-independent bias parameter.  However, our model exhibits scale dependent biasing due to the impact of halo-exclusion, the use of a ``soft'' (i.e. not infinitely sharp) halo boundary, and differences in the one halo term contributions to $\xihm$ and $\ximm$.  These features naturally lead us to a redefinition of the halo boundary that lies at the ``by eye'' transition radius from the one--halo to the two--halo term in the halo--mass correlation function. When adopting our proposed definition, our model succeeds in describing the halo--mass correlation function with $\approx 2\%$ residuals over the radial range $0.1\ \hMpc < r < 80\ \hMpc$, and for halo masses in the range $10^{13}\ \hmsun < M < 10^{15}\ \hmsun$.   Our proposed halo boundary is related to the splashback radius by a roughly constant multiplicative factor. Taking the 87-percentile as reference we find $\rt/R_{\rm sp} \approx 1.3$.  Surprisingly, our proposed definition results in halo abundances that are well described by the Press-Schechter mass function with $\dsc=1.449\pm 0.004$.  The clustering bias parameter is offset from the standard background-split prediction by $\approx 10\%-15\%$.  This level of agreement is comparable to that achieved with more standard halo definitions. 
\end{abstract}


\section{Introduction}
\label{intro}

The halo model is a powerful formalism for studying the statistical properties of the dark matter and galaxy density fields.  In the halo model, the abundance and distribution of galaxies and clusters are linked to the abundance and distribution of dark matter halos \citep{Cooray-Sheth}.  The halo model makes several key assumptions.  First, it assumes all the matter in the Universe is contained in halos.  This means that the distribution of matter in the Universe can be described by specifying the abundance and distribution of halos, as well as the mass distribution within these halos.  These statistics are described by the halo mass function $dn/dm$, the halo bias $b(m)$, and the halo density profile $u(r|m)$.  Predicting these halo properties requires large computer simulations that map the matter distribution of the Universe.  The output of the simulations is then analyzed using a halo finder.

Every halo finding algorithm makes two critical yet relatively arbitrary choices.  The first has received plenty of attention, and is the definition of halo mass.  Halo mass is typically defined as the mass enclosed within some specific spherical aperture, chosen such that the mean density of the halo within that sphere is equal to some factor of either the critical density or the mass density of the Universe.  Spherical overdensity definitions come with a number of issues, such as pseudo-evolution of halo radius and mass \citep{diemer2013a, diemer2013b}.  Recent studies have looked into more physically motivated halo boundaries, such as the splashback radius \citep{diemerkravtsov14, More2015}.  The splashback radius is defined as the radius at which accreted matter reaches its first orbital apocenter after turnaround.  This choice of radius solves the issue of pseudo-evolution and cleanly separates infalling material from matter orbiting in the halo. However, other definitions are also commonly used (e.g. friends-of-friends) \citep[see e.g.][]{Knebe2013}. For this reason, one can find calibrations of the halo mass function for multiple definitions \citep[e.g.][]{Tinker2008, Bhattacharya2011, McClintock2018}.

The second arbitrary choice is how a halo finding algorithm decides which structures are parent halos, and which are sub-halos that ``belong'' to a larger halo.  We refer to the criteria for categorizing structures as parent halos vs. sub-halos as {\it percolation} or {\it halo exclusion criteria}.  There is currently no standard percolation scheme, with different halo finders applying different halo exclusion criteria when constructing halo catalogs.  The choice of percolation can impact the halo--mass correlation function by up to $\approx 30\%$ \citep{GarciaRozo2019}.

The simplest commonly used form of a halo-model description of the halo--mass correlation function ignores both halo boundaries and halo exclusion.  One writes $\xihm (r) = \xihmone (r) + \xihmtwo (r)$ where the first and second term are referred to as the 1-halo and the 2-halo term respectively \citep[see e.g.,][]{Cooray-Sheth}. These two components are usually modeled independently.  The one--halo term is described by a halo profile $u(r|m)$, usually an NFW or Einasto profile \citep{NFW, Einasto}.  The two--halo term is modeled by assuming a scale-independent halo bias, where the bias can be defined relative to either the mass correlation function or the linear correlation function. These assumptions result in biases as large as $\approx 20\%$ at translinear scales \citep{HayashiWhite2008}. More recent efforts have introduced scale dependence of the halo bias, allowing for more accurate modeling of the trans-linear regime \citep{Surhud2013}.

In this paper, we incorporate both halo edges and halo exclusion into the modeling of the halo--mass correlation function.  We demonstrate that by explicitly introducing these two components into the model we achieve much better accuracy from small to large scales for a wide range of halo masses.  We emphasize that our model does not require any scale dependent clustering biases, beyond those brought about because of halo exclusion effects.  Our model naturally leads us to redefine halo boundaries based on the properties of the halo--mass correlation function.  In particular, we show that there is a unique halo radius and mass power-law relation $R(M)$ that ensures consistency between the halo catalog and our model.


\section{A halo model for the halo--mass correlation function}

\subsection{The Standard Approach}

We begin with a brief review of the formalism detailed in \citet{Cooray-Sheth}, as it forms the basis for our model. Let $\vec{x}_i$ be the position of the $i^{th}$ halo in the Universe.  If all mass is in contained within halos, then the mass density of the Universe can be written as
\begin{equation}
	\rho_{\rm m} (\vec{x}) = \sum_i m_i u (\vec{x} - \vec{x}_i | m_i)
\end{equation}
where $u(r|m)$ is the halo profile, and $m_i$ is the mass of the $i^{th}$ halo. Likewise, given a halo selection function $\phi(m)$ (i.e. $\phi(m)=1$ when $m \in [m-\Delta m, m+\Delta m]$ and 0 otherwise) the corresponding halo density field is 
\begin{equation}
	n (\vec{x}) = \sum_i \delta (\vec{x} - \vec{x}_i) \phi(m_i).
	\label{eq:haloden}
\end{equation}

Given these two fields, the halo--mass correlation function is
\begin{equation}
	\xihm (|\vec{x} - \vec{x}'|) = \frac{1}{\mean{n} \mean{\rho}_{\rm m}} \langle n(\vec{x}) \rho_{\rm m} (\vec{x}') \rangle - 1
	\label{hmcf}
\end{equation}

where $\langle \cdots \rangle$ denotes ensemble averaging. We can plug in the expressions for each density field into \ref{hmcf}, and predict the two-point correlation function in therms of the halo density profile, the halo mass function, and the clustering of halos. One has then
\begin{align}
	\langle n(\vec{x}) \rho_{\rm m} (\vec{x}') \rangle &= \Big \langle \sum_i \sum_j m_j \phi (m_i) \delta (\vec{x} - \vec{x}_i) u (\vec{x}' - \vec{x}_j | m_j) \Big \rangle \nonumber \\
    											 &= \Big \langle \sum_i m_i \phi (m_i) \delta (\vec{x} - \vec{x}_i) u (\vec{x}' - \vec{x}_i | m_i) \Big \rangle \nonumber \\
    											 &+ \Big \langle \sum_i \sum_{j \neq i} m_j \phi (m_i) \delta (\vec{x} - \vec{x}_i) u (\vec{x}' - \vec{x}_j | m_j) \Big \rangle
\end{align}

The average over the ensemble has been separated into two parts: one that accounts for the correlation between a halo and the mass contained within it, and one that accounts for the correlation between a halo, and mass that belongs to other halos. We treat each in turn.  We have 
\begin{align}
	1^{st}\ term &= \Big \langle \sum_i m_i \phi (m_i) \delta (\vec{x} - \vec{x}_i) u (\vec{x}' - \vec{x}_i | m_i) \Big \rangle \nonumber \\
        		 & \hspace{-0.5in} = \Big \langle \int dm\ \sum_i m_i \phi (m_i) \delta (\vec{x} - \vec{x}_i) u (\vec{x}' - \vec{x}_i | m_i) \delta (m - m_i) \Big \rangle \nonumber \\
                 & \hspace{-0.5in} = \int dm\ m \phi (m) u (\vec{x}' - \vec{x} | m) \Big \langle \sum_i \delta (\vec{x} - \vec{x}_i) \delta (m - m_i) \Big \rangle
\end{align}

The remaining expectation value corresponds to the mean number of halos per unit volume per unit mass, that is, the halo mass function,
\begin{equation}
	\frac{dn}{dm} = \Big \langle \sum_i \delta (\vec{x} - \vec{x}_i) \delta (m - m_i) \Big \rangle
\end{equation}
Plugging the mass function into the $1^{st}\ term$ and integrating over a narrow mass selection function we arrive at
\begin{align}
	1^{st}\ term &= \int dm\ \frac{dn}{dm} m \phi (m) u (\vec{x}' - \vec{x} | m) \nonumber \\
	             &= \mean{n} m u (\vec{x}' - \vec{x} | m)
\end{align}
This is the so-called one halo term of the halo--mass correlation function.

Now, let's look at the second term,
\begin{align}
	2^{nd}\ term &= \Big \langle \sum_i \sum_{j \neq i} m_j \phi (m_i) \delta (\vec{x} - \vec{x}_i) u (\vec{x}' - \vec{x}_j | m_j) \Big \rangle \nonumber \\
                 & \hspace{-0.3in} = \int dm dm' d\tilde{\vec{x}}\ m' \phi (m) u (\vec{x}' - \tilde{\vec{x}} | m') \nonumber \\
                 & \hspace{-0.3in} \times \Big \langle \sum_i \sum_{j \neq i} \delta (\vec{x} - \vec{x}_i) \delta(m - m_i) \delta(\tilde{\vec{x}} - \vec{x}_j) \delta(m' - m_j) \Big \rangle \nonumber \\
                 & \hspace{-0.3in} = \int dm dm' d\tilde{\vec{x}}\ m' \phi (m) u (\vec{x}' - \tilde{\vec{x}} | m') \nonumber \\
                 & \hspace{-0.3in} \times \frac{dn}{dm} \frac{dn}{dm'} [1 + \xihh(\vec{x} - \tilde{\vec{x}} | m, m')]
                 \label{eq:2ndterm}
\end{align}

Halos are biased tracer of the matter density field.  At scales much larger than the size of halos $\xihh (r|m,m') = b(m) b(m') \xi_L (r)$
\begin{equation}
	\xihh(\vec{x} - \tilde{\vec{x}} | m, m') = b(m) b(m') \xi_L (\vec{x} - \tilde{\vec{x}})
\end{equation}

It follows that
\begin{align}
	2^{nd}\ term &= \mean{n} \mean{\rho}_{\rm m} + \mean{n} b(m) \int dm'\ \frac{dn}{dm'} m' b(m') \nonumber \\
                 &\times \int d\tilde{\vec{x}}\ u (\vec{x}' - \tilde{\vec{x}} | m') \xi_L (\vec{x} - \tilde{\vec{x}})
\end{align}

At large scales, the details of the halo profile become unimportant, and the halos themselves can be approximated as point masses, so that $u(\vec x) \approx \delta(\vec x)$.  With this approximation, and the identity,
\begin{align}
    \int dm' \frac{dn}{dm'} m' b(m') = 1\,,
\end{align}
the $2^{nd}$ term becomes
\begin{align}
	2^{nd}\ term &= \mean{n} \mean{\rho}_{\rm m} + \mean{n} \mean{\rho}_{\rm m} b(m) \xi_L (\vec{x} - \vec{x}')\,.
\end{align}

Getting everything together, the product becomes
\begin{align}
	\langle n(\vec{x}) \rho_{\rm m} (\vec{x}') \rangle &= \mean{n} m u (\vec{x}' - \vec{x} | m) + \mean{n} \mean{\rho}_{\rm m} (1 + b(m) \xi_L (\vec{x} - \vec{x}'))\,.
\end{align}

The halo mass correlation function is
\begin{align}
	\xihm (r | m) &= \frac{m}{\mean{\rho}_{\rm m}} u (r | m) + b(m) \xi_L (r),
\end{align}
where $r = |\vec{x} - \vec{x}'|$. The first term is known as the one-halo term, $\xihmone$, and accounts for the mass within a single halo. The second term is known as the two-halo term, $\xihmtwo$, and accounts for the mass across different halos.


\subsection{Incorporating Halo Exclusion}

In the standard approach, we assumed that
\begin{align}
	\Big \langle \sum_i \sum_{j \neq i} \delta (\vec{x} - \vec{x}_i) \delta(m - m_i) \delta(\tilde{\vec{x}} - \vec{x}_j) \delta(m' - m_j) \Big \rangle \nonumber \\
	= \frac{dn}{dm} \frac{dn}{dm'} [1 + b(m) b(m') \xi_L (\vec{x} - \tilde{\vec{x}})]
\end{align}

This is true at large scales, because halos never overlap. This is not the case at small scales. We introduce a halo exclusion function $E(\vec{x}_i - \vec{x}_j | m_i, m_j)$ which is zero when halos overlap, and one otherwise.  This halo exclusion function multiplies the entire $2^{nd}$ term, so that equation~\ref{eq:2ndterm} now becomes
\begin{align}
	2^{nd} &= \int dm dm' d\tilde{\vec{x}}\ m' \phi (m) u (\vec{x}' - \tilde{\vec{x}} | m') \nonumber \\
	       &\times \frac{dn}{dm} \frac{dn}{dm'} [1 + b(m) b(m') \xi_L (\vec{x} - \tilde{\vec{x}})] E(\vec{x} - \tilde{\vec{x}} | m, m')
\end{align}

For a narrow selection function, we get
\begin{align}
	2^{nd} &= \int dm'\ \mean{n} \frac{dn}{dm'} m' \int d\tilde{\vec{x}}\ u (\vec{x}' - \tilde{\vec{x}} | m') \nonumber \\
	       &\times [1 + b(m) b(m') \xi_L (\vec{x} - \tilde{\vec{x}})] E(\vec{x} - \tilde{\vec{x}} | m, m')
\end{align}

The integral over all space is a convolution of the density profile and the 2-halo term with exclusion.
\begin{align}
	2^{nd} &= \int dm'\ \mean{n} \frac{dn}{dm'} m'  (u \ast E) (\vec{x} - \vec{x}' | m, m') \nonumber \\
	       &+ \int dm'\ \mean{n} \frac{dn}{dm'} m' b(m) b(m')  (u \ast E \xi_L) (\vec{x} - \tilde{\vec{x}})
\end{align}

To make further progress, we must specify a halo exclusion function.  We assume halo exclusion happens when halos are separated by a distance $r\leq \re (m, m')$, where $\re$ is the halo exclusion radius. Note that the halo exclusion radius depends on the masses $m$ and $m'$ of the two halos under consideration.  With this definition, the halo exclusion function takes the form
\begin{equation}
	E(r | m, m') = 1 - \theta(\re (m, m') - r)
    \label{exclusion}
\end{equation}
where $\theta$ is the Heaviside step function.  We can set upper and lower bounds for the exclusion radius. For the lower bound, the exclusion radius must be larger than the radius of either of the two halos. For the upper bound, we use a hard sphere model.
\begin{equation}
	\max \{\rt(m), \rt(m')\} < \re (m, m') < \rt(m) + \rt(m')
    \label{ineq}
\end{equation}
In the above expression, $\rt(m)$ is the radius of a halo of mass $m$.

Inserting \ref{exclusion} into our previous expressions we find
\begin{align}
	2^{nd} &= \int dm'\ \mean{n} \frac{dn}{dm'} m'  [u \ast (1 - \theta_e)] (r) \nonumber \\
	       &+ \int dm'\ \mean{n} \frac{dn}{dm'} m' b(m) b(m')  [u \ast (1 - \theta_e) \xi_L] (r) \\
	       &= \mean{n} \mean{\rho}_{\rm m} \left[1 - \int dm'\ \frac{dn}{dm'} \frac{m'}{\mean{\rho}_{\rm m}}  (u \ast \theta_e) (r) \right] \nonumber \\
	       &+ \mean{n} \mean{\rho}_{\rm m} b(m) \left[ \int dm'\ \frac{dn}{dm'} \frac{m'}{\mean{\rho}_{\rm m}} b(m')  [u \ast (1 - \theta_e) \xi_L] (r) \right]
\end{align}

The halo--mass correlation function becomes
\begin{align}
	\xihm(r | m) &= \frac{m}{\mean{\rho}_{\rm m}} u(r | m) - \int dm'\ \frac{dn}{dm'} \frac{m'}{\mean{\rho}_{\rm m}} \theta_e(r | m, m') \nonumber \\
    		     &+ b(m) \int dm'\ \frac{dn}{dm'} \frac{m'}{\mean{\rho}_{\rm m}} b(m')  [u \ast (1 - \theta_e) \xi_L] (r)
\end{align}

Note that this model for the halo--mass correlation function explicitly incorporates halo exclusion in a flexible way, in the sense that the model can be used with any definition for a halo boundary and with any choice of halo percolation.

We can further simplify this expression by using the same approximation as in the standard case, i.e. at large scales the mass profile becomes unimportant, and we can set $u(\vec x) \approx \delta(\vec x)$. With this approximation, the above expression simplifies to
\begin{align}
	\xihm(r | m) &= \frac{m}{\mean{\rho}_{\rm m}} u(r | m) +b(m) \xi_L (r) \nonumber \\
	             &- \int dm'\ \frac{dn}{dm'} \frac{m'}{\mean{\rho}_{\rm m}} \theta_e(r | m, m') \nonumber \\
    		     &- b(m) \xi_L (r) \int dm'\ \frac{dn}{dm'} \frac{m'}{\mean{\rho}_{\rm m}} b(m') \theta_e (r | m, m') .
\end{align}
%

We wish to incorporate into our model the fact that spherical-overdensity halo finders define sharp halo edges such that the mass interior to the halo radius belongs to the halo, while mass exterior to the halo radius does not.   This in turn implies that a self-consistent model of the halo--mass correlation function ought to truncate the halo term at the halo boundary.  With this truncation in mind, the matter density field can be written as
\begin{align}
	\rho_{\rm m} (\vec{x}) = \sum_i m_i u (\vec{x} - \vec{x}_i | m_i) \theta (r | \rt(m))
\end{align}
where $\theta (r | \rt(m)) = 1$ when $r < \rt(m)$ and $0$ otherwise. This imposes a sharp cut in the halo density profile of one halo.  In practice, however, we expect that the halo--mass correlation function will exhibit some effective finite width in the radial direction.  For instance, we know halos are triaxial, so even if a halo is defined using a spherical overdensity, we expect ``nature'' would prefer a triaxial definition.  A triaxial halo definition would ``spread out'' the halo boundary across a range of radial scales, naturally leading to a soft truncation of the one-halo term.  In short, we expect a soft truncation will produce better results than an infinitely sharp truncation.  Of course, this implies that our model is not entirely consistent with the sharp radial cut imposed by halo finders.  We consider this a small price to pay for better precision in our model.  Moreover, one could imagine modifying halo finders in order to implement a soft truncation, thereby mimicking our model for the halo--mass correlation function.  Indeed, this is how some cluster finders work \citep[e.g. redMaPPer][]{rykoffetal12}.  We will leave the task of exploring such modifications of halo finders to future work.

For the above reasons, we choose to model the truncation using the complementary error function centered at the halo edge $\rt(m)$ with width $\Delta \rt(m)$.
\begin{align}
    \thetat(r | m) = \frac{1}{2} \erfc{\left( \frac{r - \rt}{\sqrt{2}\Delta \rt} \right)}
    \label{eq:smooth_truncation}
\end{align}

Upon including this halo truncation term, the halo--mass correlation function can be written as
\begin{align}
	\xihm(r | m) &= \frac{m}{\mean{\rho}_{\rm m}} u(r | m) \thetat(r | m) + b(m) \xi_L (r) \nonumber \\
    				&- \int dm'\ \frac{dn}{dm'} \frac{m'}{\mean{\rho}_{\rm m}} \theta(r | \re(m, m')) \nonumber \\
    				&- b(m) \xi_L (r) \int dm'\ \frac{dn}{dm'} \frac{m'}{\mean{\rho}_{\rm m}} b(m') \theta(r | \re(m, m'))
\end{align}

This final expression can still be interpreted in a similar fashion as the standard halo model. It has a one--halo term that accounts for the matter that is contained within the halo boundary, and has a two--halo term that takes into account matter in the rest of the halos. The difference is that there are correction terms due to halo exclusion. These correction terms can be interpreted as the mass that would have been there in other halos, were it not for the exclusion volume associated with more massive halos. Note this ``excluded mass'' is comprised of both the excluded mass in the mean, and the ``extra'' excluded mass due to halo--mass clustering.  The final expressions for the 1-halo and 2-halo terms are
\begin{align}
	\xihmone (r | m) &= \frac{m}{\mean{\rho}_{\rm m}} u(r | m) \thetat(r | m) - \int dm'\ \frac{dn}{dm'} \frac{m'}{\mean{\rho}_{\rm m}} \theta(r | \re(m, m')) \nonumber \\
    				&- b(m) \xi_L (r) \int dm'\ \frac{dn}{dm'} \frac{m'}{\mean{\rho}_{\rm m}} b(m') \theta(r | \re(m, m'))
\end{align}
\begin{align}
	\xihmtwo (r | m) &= b(m) \xi_L (r) .
\end{align}

Note we have associated the correction terms with the one-halo term since these represent excluded mass in the vicinity of the halo, i.e. the excluded mass moves in space as one moves halos in space.

\subsection{Refining the Two--halo Term}

In all of the above we have assumed that $\xihh = b(m) b(m') \xi_L$.  This is true at very large scales but not at small scales.  If halos trace matter, then as we move towards non-linear scales, we should expect $\xi_L$ will need to be replaced the matter--matter correlation function $\ximm$.  However, the latter correlation function has a strong 1-halo contribution at small scales.  Clearly, linear biasing can't hold in this regime.  The best we could hope for is linear bias relative to the 2-halo term of the matter correlation function, $\ximm^{\rm 2h}=\ximm - \ximm^{\rm 1h}$.  This raises the obvious question: how can we remove the 1-halo term of the matter correlation function? 

While we cannot give a definitive answer a priori, it is clear what ``removing the 1-halo term'' must do to the matter correlation function: it must suppress correlations at small scales.  This leads us to adopt a two-halo term for the matter--matter correlation function of the form given by 
\begin{align}
    \ximmtwo &= \ximm \times (1 - \thetat (\reff, \Deltaeff)) .
\end{align}
In this expression, $\thetat$ is again a smooth truncation function of the form given by equation~\ref{eq:smooth_truncation}. The radius $\reff$ sets the scale at which $\ximm$ transitions from the 1-halo term to the 2-halo term, while $\Deltaeff$ determines how quickly this transition occurs. 

With these modifications, our final expression for the halo--matter correlation function is
\begin{align}
	\xihm(r | m) &= \xihmone(r|m) + \xihmtwo(r|m) \label{hmcf-final} \\
	\xihmone(r|m) &= \frac{m}{\mean{\rho}_{\rm m}} u(r | m) \thetat(r | m) \nonumber \\
	            &- \int dm'\ \frac{dn}{dm'} \frac{m'}{\mean{\rho}_{\rm m}} \theta(r | \re(m, m')) \nonumber \\
    				&- b(m) \ximmtwo (r) \int dm'\ \frac{dn}{dm'} \frac{m'}{\mean{\rho}_{\rm m}} b(m') \theta(r | \re(m, m')) \label{hmcf-final1}\\
    \xihmtwo(r|m) &= b(m) \ximmtwo (r) \label{hmcf-final2} 
\end{align}

We briefly discuss how the expressions written above compare with the expressions for the halo matter correlation function in \citet{Surhud2013}. In their approach, the first term in eq.~\ref{hmcf-final1} is $\xihmone$, while the rest of the terms in that equation are accounted for in their two halo term.  The two halo term that they consider includes radial dependence of the halo bias, halo exclusion, and uses the non-linear matter correlation function. In our case, we consider a simple linear bias relative to $\ximmtwo$ instead. Thus there is considerable simplicity in the expressions we have derived. In Section~\ref{sec:results}, we will fit the halo matter correlation function with the results from numerical simulations. For the routinely used halo mass definition $M_{\rm 200m}$, the model of \citet{Surhud2013} performs well while our model performs poorly.  As we show below, however, if the halo definitions are made consistent with our formalism --- a step that requires fairly simple and straightforward tweaks to the halo finding algorithms --- the simpler expressions in our model can describe the halo mass correlation function with even greater accuracy than that achieved by \citet{Surhud2013}.  In particular, we argue that the complications regarding the radial dependence of the halo bias can be solved by a simple redefinition of the halo boundary, coupled with the use of the 2-halo term of the non-linear matter correlation function.

\subsection{High mass limit}

The most massive halos are much bigger than the rest of their neighbors ($m >> m'$). Consequently, $\rt(m) >> \rt(m')$. This condition, along with the inequality \ref{ineq} implies that,
\begin{align}
	\re(m, m') &\approx \rt(m)
\end{align}

Setting the exclusion radius to the halo radius of the more massive halo leads to
\begin{align}
	\xihm (r | m) &= \left[ \frac{m}{\mean{\rho}_{\rm m}} u(r | m) - 1 \right] \thetat(r | m) \nonumber \\
	&+ b(m) \ximmtwo (r) [1 - \thetat(r | m)]
\end{align}

which is equivalent to
\begin{align}
   \xihm (r | m) =
   \begin{cases} 
      \frac{m}{\mean{\rho}_{\rm m}} u(r | m) - 1 & r\leq \rt(m) \\
      b(m) \ximmtwo (r) & r\geq \rt(m)
   \end{cases}
\end{align}

If instead of the truncated matter--matter correlation function we use the linear correlation function $\xi_L$ the model turns to
\begin{align}
   \xihm (r | m) =
   \begin{cases} 
      \frac{m}{\mean{\rho}_{\rm m}} u(r | m) - 1 & r\leq \rt(m) \\
      b(m) \xi_L (r) & r\geq \rt(m)
   \end{cases}
\end{align}
Note that if $\rt(m)$ is not known a priori, one can use the fact that $\xihm$ is continuous to determine $\rt(m)$.  That is, our formalism has allowed us to derive from first principles the model proposed by \cite{HayashiWhite2008}.

\subsection{Halo density profile}
We assume that dark matter halos are spheres whose normalized density distribution is given by the Einasto profile
\begin{align}
u(r | m) &= \frac{\rhos}{m} \exp \left\{ -\frac{2}{\alpha} \left[ \left( \frac{r}{\rs} \right)^{\alpha} - 1 \right] \right\}
\end{align}

where $\rs$ is the scale radius, $\rhos$ the density at $\rs$ and $\alpha$ is the shape parameter.  In the following we use a more convenient parameterization via the mass and concentration and a mass definition.  For a particular mass definition, say $M_{\rm 200m}$, the concentration is given by $c = R_{\rm 200m} / \rs$, and the density $\rhos$ is obtained by the normalization of $u(r | m)$.  This parameterization is
\begin{align}
u(r | M) &= \frac{\alpha c^3 (\frac{2}{\alpha})^{3/\alpha}}{3 \Gamma(3/\alpha) M} \exp \left\{ -\frac{2}{\alpha} \left( \frac{r c}{R(M)} \right)^{\alpha} \right\}
\end{align}
where $R(M)$ is the mass contained within the radius $R$ that defines the boundary of the halo. We use the {\it colossus} \citep{colossus} python package to compute the Einasto profile.

\subsection{Simplifying the halo exclusion terms}

The fact that the exclusion terms are integrals over an infinite mass range poses a problem.  With the simulation we are using we simply don't know the halo mass function for $M < 10^{12} \hmsun$.  Moreover, lowering this mass limit requires using smaller boxes, which in turn looses larger modes. In short, a  brute-force force approach to this problem appears unpalatable.  However, we can take advantage of the fact that the corrections due to halo exclusion are integrals over the halo mass function.  Adopting the normalization condition that all mass is contained within halos, we can think of the mass function $(1/\bar \rho_{\rm m}) m dn/dm$ as a probability distribution.  Consequently, the mass integrals can all be thought of as expectation values.  We assume that the average value of a function $f$ over the probability distribution $(1/\bar \rho_{\rm m}) m dn/dm$  can be approximated as the function $f$ evaluated at some input parameter $m_0$ where we expect $m_0 \approx \avg{m}$.

Using this approximation, the corrections due to halo exclusion are simplified and the correlation function can be written as
\begin{align}
	\xihm(r | m) &= \frac{m}{\mean{\rho}_{\rm m}} u(r | m) \thetat(r | m) + b(m) \ximmtwo (r) \nonumber \\
    				&- \theta(r | \re(m, m_b)) - b(m) \ximmtwo (r) \theta(r | \re(m_a, m'))
\end{align}
where $m_a$ and $m_b$ are the values of halo mass which approximate the expectation value of the exclusion functions over all halo masses.  There are two parameters since the exclusion function is weighted differently in each exclusion term. Note in particular that the parameter $m_a$ arises from the exclusion function integral that includes a bias weighting of the halos, so we should expect $m_a>m_b$ due to the steep dependence of the halo bias at high masses.

Our approach here again differs from that of \citet{Surhud2013}, who split the mass integrals into two, and then simplified using the integral conditions.  Nevertheless, they must still perform integrals over mass, which our model does away through the introduction of the $m_a$ and $m_b$ parameters.

\subsection{Model parameters}

Our final model for the halo--mass correlation function depends on several model parameters, namely:
\begin{itemize}
    \item concentration $c$
    \item Einasto parameter $\alpha$
    \item halo bias $b$
    \item truncation parameters of $\ximmtwo$: $\reff$, $\Deltaeff$
    \item effective masses for halo exclusion corrections $m_a$, $m_b$
\end{itemize}

There are additional ``parameters'' in our fits, namely
\begin{itemize}
    \item the halo mass $m$,
    \item the halo radius $\rt$.
\end{itemize}
The mass $m$ governs the amplitude of the 1-halo term in our fit, while the radius $\rt$ sets the boundary of the halo. In principle, these parameters should \it not \rm be fit parameters.  For instance, when using an overdensity criterion $\Delta$ when defining halo masses, a self-consistent model should have $M_\Delta$ as the mass parameter governing the 1-halo amplitude.  Likewise, one should set the radius $\rt=R_\Delta$.

As we will see, in practice, using commonly-used fixed overdensity criteria results in poor fits to the data.  This allows us to ask the question: is the simulation data well fit with some other halo mass $m$ and halo radius $\rt$?  In this case, we can use our halo model with $m$ and $\rt$ as fit parameters to learn about what the mass and radius of the halos should have been.  When doing so, our fits rely on 9 parameters for a single mass bin.  However, we can vastly reduce this parameter space by enforcing simple power-law scalings of many of our parameters with halo mass. Additionally, the one--halo term that we subtract from the matter--matter correlation function must be independent of mass. Thus, the parameters $\reff$ and $\Deltaeff$ have to be shared across all halo mass bins. This forces us to simultaneously fit the model across all available halo masses simultaneously.

We assume that the halo radius, concentration, the shape parameter, and the effective masses can be parameterized as power laws of halo mass. That is
\begin{align}
    \rt &= r_p \left( \frac{m}{m_{p1}} \right)^{\beta}
\end{align}
\begin{align}
    c &= c_p \left( \frac{m}{m_{p2}} \right)^{\gamma}
\end{align}
\begin{align}
    \alpha &= \alpha_p \left( \frac{m}{m_{p3}} \right)^{\delta}
\end{align}
\begin{align}
    m_a &= m_{ap} \left( \frac{m}{m_{p4}} \right)^A
\end{align}
\begin{align}
    m_b &= m_{bp} \left( \frac{m}{m_{p5}} \right)^B
\end{align}

where we fit for the amplitudes and exponents in these relations.  We select the pivot values $m_{p1} = m_{p3} = 2\times10^{14}$, $m_{p2}= 7\times10^{14}$, $m_{p4}=m_{p5} = 2\times10^{12}$ which are typical values of halo mass in our halo catalog.  The above selection of pivot points roughly decorrelates the slope and amplitude parameters, and was obtained through  trial and error.

The likelihood of the halo--mass correlation function for halos in the $k{\rm th}$ mass bin is
\begin{align}
    \ln \mathcal{L}_k = \ln \mathcal{L} (\xihm  (r | m_{\rm k}) | \theta) &\propto - \frac{1}{2} \mathbf{D}_k^\top \mathbf{C}_{\xihm}^{-1} \mathbf{D}_k
\end{align}
where $\theta = (m, b, r_{\rm p}, \beta, c_{\rm p}, \gamma, \alpha_{\rm p}, \delta, m_{\rm ap}, A, m_{\rm bp}, B, \reff, \Deltaeff)$ is the vector of model parameters, $\mathbf{D}_k = \xihm^{\rm data} - \xihm^{\rm model}$ and $\mathbf{C}_{\xihm}$ is the covariance matrix of $\xihm^{\rm data}$.  We are looking to fit for all mass bins simultaneously since the parameters $\reff$ and $\Deltaeff$ are shared across all halo mass bins.  To maintain the jackknife covariance matrix well-conditioned, we ignore the covariance across mass bins.  We emphasize that while this assumption will impact the width of the posterior distribution in our analysis, we expect its impact on the precision of the best fit model will be minimal.  With this assumption, the total likelihood is given by
\begin{align}
    \ln \mathcal{L} (\{\xihm(r | m_k)\}_{k=1}^N | \mathbf{\theta)} &\propto \sum_k \ln \mathcal{L}_k
\end{align}

The priors on the parameters are shown in table \ref{table:parameters}.  The likelihood is sampled using the python package {\it emcee} \citep{emcee}.  The total number of parameters for 12 mass bins is 38 (i.e. just over 3 parameters per correlation function).  We use 152 walkers with 50000 steps each and discard the first 5000 steps of each walker.  The chains of each walker become uncorrelated after 400 steps, ensuring a minimum of 17,000 independent samples.

\begin{table}
\begin{tabular}{|l|l|l|}
Parameter & Description & Prior \\
\hline
$\log_{10} m_k$ & Halo mass & $[11.0, 16.0]$ \\ 
$b_k$ & Halo bias & $[0, \infty]$ \\
$r_{\rm p}$ & Halo radius pivot & $[0, \infty]$ \\
$\beta$ & Halo radius power & $[0, \infty]$ \\
$c_{\rm p}$ & Concentration pivot & $[0, \infty]$ \\
$\gamma$ & Concentration power & $[-\infty, 0]$ \\
$\alpha_{\rm p}$ & Shape parameter pivot & $[-\infty, \infty]$ \\
$\delta$ & Shape parameter power & $[0, \infty]$ \\
$m_{\rm ap}$ & Halo exclusion pivot & $[0, \infty]$ \\
$A$ & Halo exclusion power & $[0, \infty]$ \\
$m_{\rm bp}$ & Halo exclusion pivot & $[0, \infty]$ \\
$B$ & Halo exclusion power & $[0, \infty]$ \\
$\Delta$ & Width of truncation for one--halo term & $[0, \infty]$ \\
$\Delta_e$ & Width of truncation for exclusion terms & $[0, \infty]$ \\
$\reff$ & Truncation radius of $\ximm$ & $[0, \infty]$ \\
$\Deltaeff$ & Truncation width of $\ximm$ & $[0, \infty]$ \\
\end{tabular}
\label{table:parameters}
\caption{Model parameters.  The limits in square brackets indicate flat priors.}
\end{table}

\section{Results}
\label{sec:results}
\subsection{The Halo--Mass Correlation Function}

We measure the halo--mass correlation function using a cosmological N-body simulation similar to those used in the Aemulus project \citep{deroseetal2018}.  It is run with the publicly available code \verb+GADGET2+ \citep{Springel2005}.  The simulation is a periodic box of size $1050\ \hMpc$ with $1400^3$ particles.  The cosmology is $h=0.6704$, $\Omega_m=0.318$, $\Omega_{\Lambda}=0.682$, $\Omega_b=0.049$, $\sigma_8=0.835$, $n_s=0.962$.  The particle mass is $3.7275 \times 10^{10} \ \hmsun$ and the force softening scale is $20\ h^{-1} {\rm kpc}$.  Halos were found using the publicly available \rockstar\ halo finder \citep{Behroozi2013} with and spherical overdensity of $\Delta=200$.  \rockstar\ uses an adaptive friends-of-friends algorithm in 6-dimensional phase space to identify dark matter structures.  These structures are classified as parent halos or subhalos using a soft-sphere halo exclusion scheme: two structures are considered to be in the same parent halo if their separation is less than the radius of the larger structure.

\begin{figure*}
\includegraphics[width=80mm]{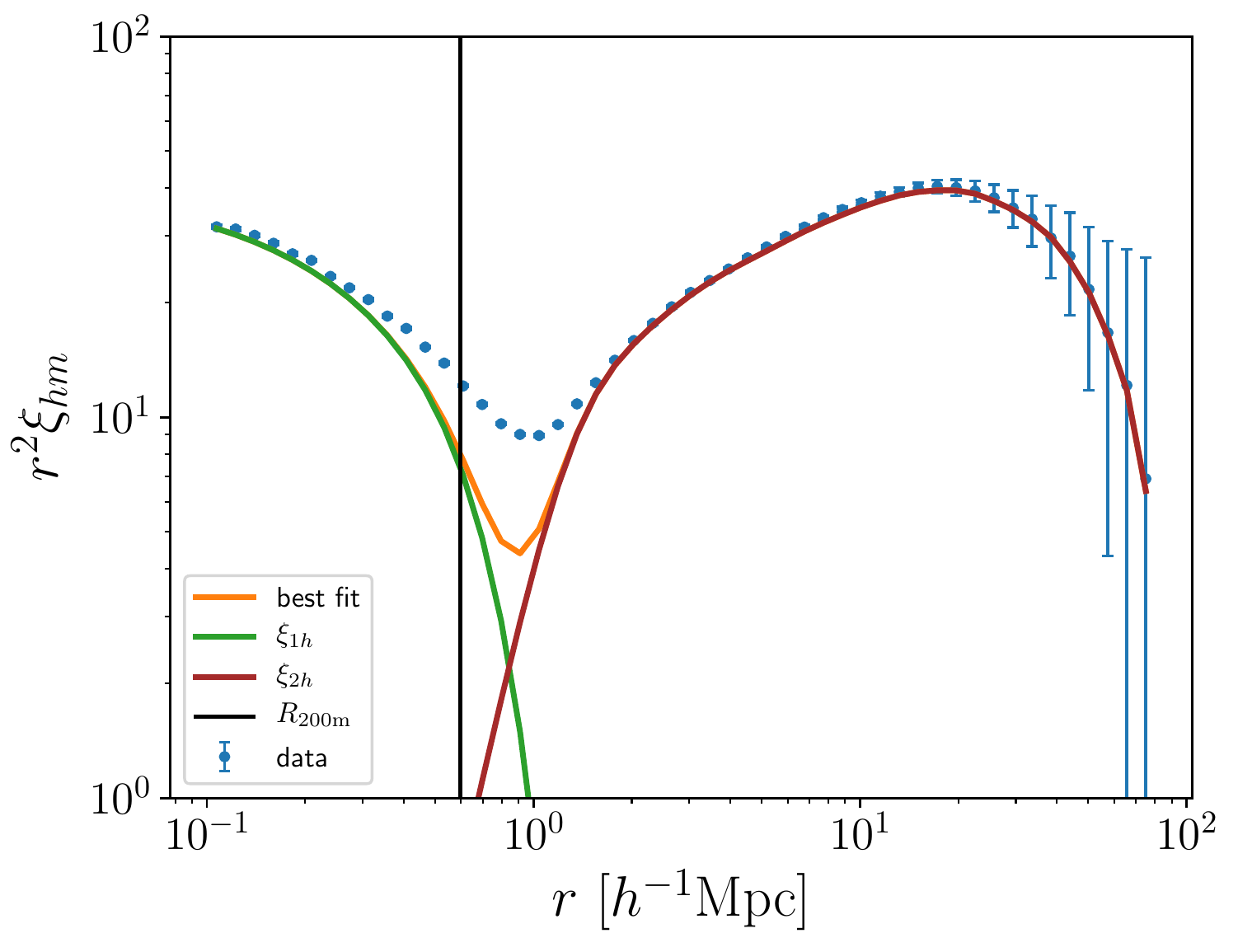} \includegraphics[width=80mm]{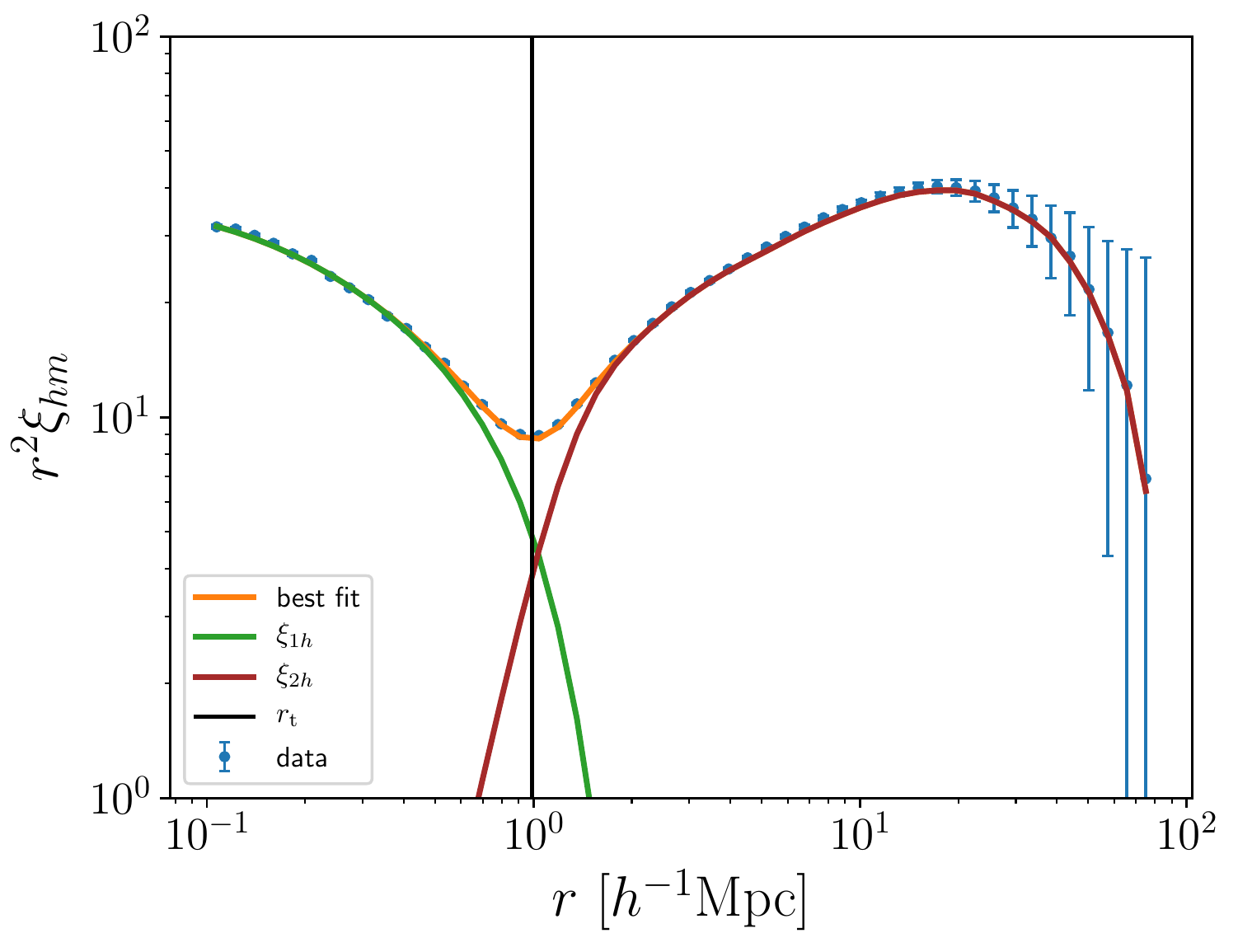}
\caption{The halo--mass correlation function and model predictions using different choices of halo radius, and letting the amplitude of the one-halo term be given by the mass contained within the appearture used to define halos. \textbf{Left}: Best fit model using $R_{\rm 200m}$ as the halo boundary and exclusion radius. \textbf{Right:} Best fit model using our self-consistent halo boundary $\rt$ obtained through iteratively applying our model to the simulations. Error bars are jackknife.  Our halo model provides an accurate description of the simulation data provided the halo boundary is properly defined.}
\label{fig:r200m-rt}
\end{figure*}

We attempt to fit the halo--mass correlation function data with our halo model.  The left panel of Figure~\ref{fig:r200m-rt} shows the halo--mass correlation function for halos of mass $M=[1, 2]\times10^{13}\ \hmsun$, where mass is defined using an overdensity criterion $\Delta=200$ relative to the mean density of the Universe.  In this fit, we have forced the mass parameter in our halo model to be equal to the mass of the halos.  Likewise, we have forced the truncation parameter $\rt$ to coincide with $R_{\rm 200m}$, the radius of the halos.  The latter is shown as a vertical line in the plot, which is left of the ``valley'' between the two bumps in the data, which one might expect to correspond to the one and two halo terms.  Unsurprisingly, the fit to the data is poor despite the model having 7 free parameters.

We now test how well our model works if we let the mass and radius parameters in the halo model be free.  This, of course, results in a model that is inconsistent with the halo definitions employed in the creation of the halo catalog.  We will address this point momentarily.  For now, let us simply consider how our model fits the data when we let $m$ and $\rt$ float.

The right panel in Figure~\ref{fig:r200m-rt} shows our best fit model for the halo--mass correlation function when allowing the mass and truncation radius parameters to float.  We see that our halo model now provides an excellent description of the data, and that the best fit truncation radius $\rt$ (shown as a vertical line) falls close to the ``by-eye'' transition between the 1- and 2-halo bumps of the halo--mass correlation function.  In other words, the simulation data clearly suggests that the halo boundary should extend further out than $R_{\rm 200m}$, and should be set by $\rt$ instead.

These results suggest how to address the lack of consistency between the model parameters $m$ and $\rt$, and the mass and halo boundary used to define the halos in the first place.  We consider an iterative approach to halo finding which proceeds as follows.  We start by assuming that halos are defined in iteriation $i$ via a radius--mass relation $R_i(M)$.  For instance, in iteration $i=1$, this relation corresponds to the fixed overdensity criterion, $3M/4\pi R_1(M)^3 = \Delta \bar \rho_{\rm m}$.  Given the relation $R_i(M)$, we perform the following operations:
\begin{enumerate}
    \item We generate a halo catalog using $R_i(M)$ to define the boundaries of halos used to enforce halo exclusion.
    \item We measure the halo--matter correlation functions for halos in fixed mass bins.
    \item We fit the resulting halo--mass correlation function letting mass $m$ and radius $\rt$ parameters float.  These new estimates define the radius--mass relation $R_{i+1}(M)$.
\end{enumerate}
The procedure is then iterated until convergence is achieved, that is, we iterate until $R_{i+1}(M)=R_i(M)$.  In practice, we find that $R(M)$ converges to within $\approx 1\%$ by the end of the second iteration, and it is converged to $0.01\%$ in $\approx 5$ iterations.  Our fully converged radius--mass relation takes the form
\begin{align}
    \rt(m) = r_p \left( \frac{m}{m_p} \right) ^ \beta
    \label{eq:rt(m)}
\end{align}

where we select the pivot value $m_p = 2 \times 10^{14} \hmsun$, and the converged parameters are $r_p = 1.558 \pm 0.001 \hMpc$, $\beta = 0.200 \pm 0.001$. This relation can be recast as a mass-dependent overdensity criterion, 
\begin{align}
    \Delta(M) = \frac{3 m_p^{3\beta }}{4\pi \rho r_p^3} M^{1-3\beta}
\end{align}

We emphasize that once convergence is achieved, the halo mass and radius should no longer be considered fit parameters.  That is, when adopting the halo mass definition described in equation \ref{eq:rt(m)}, the parameters $m$ and $\rt$ in the halo model are given precisely by the mass and radius used to define the halos.

\begin{figure*}
\hspace{-12pt} \includegraphics[width=180mm]{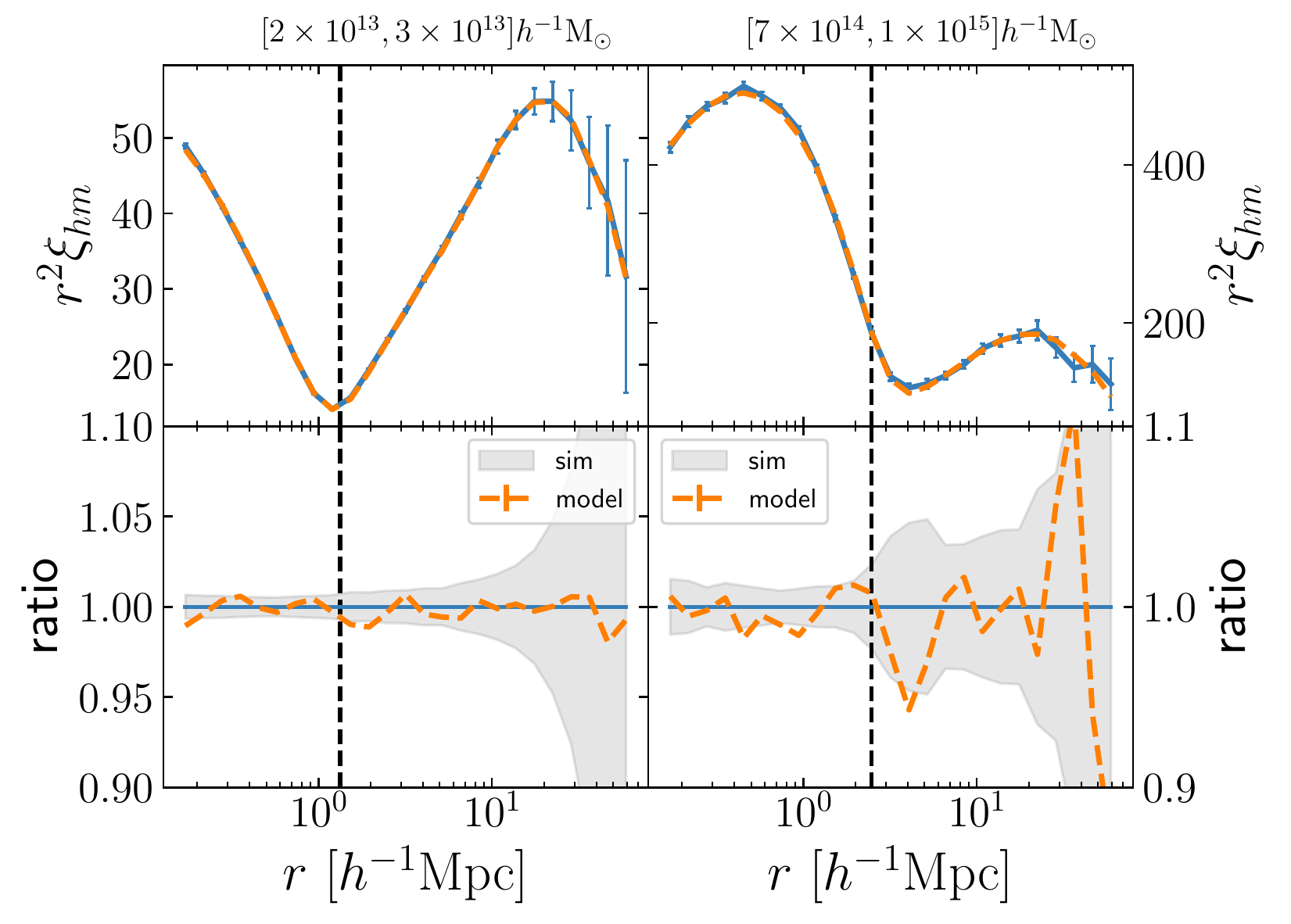}
\caption{The halo--mass correlation function for halos of different masses, as labelled. \textbf{Top row:} Halo--matter correlation functions. \textbf{Bottom row:} Fractional difference between the model and the measurement. Error bars are jackknife.}
\label{fig:hmcf}
\end{figure*}

Figure \ref{fig:hmcf} shows a comparison between the halo--matter correlations measured in the simulation to our best fit model after convergence is achieved.  The model performs well in a wide range of halo masses and scales.  It achieves 2\% accuracy for halos of mass $10^{13}\ \hmsun$ from $0.2\ \hMpc$ to $60\ \hMpc$.  Larger halo masses exhibit larger ($\sim 10\%$) deviations, though these are consistent with noise as estimated using jackknife resampling.  In other words, our simulation box is not sufficiently large for us to give a robust estimate of the precision of our model at high halo masses.  Likewise, the Press-Schechter fit presented here has only been validated for halos with mass $M\geq 10^{13}\ M_\odot$.   We will provide improved calibrations of the precision of our model in future work.

\subsection{Is $\mathbf{r}_{\mathbf{\rm t}}$ Related to the Splashback Radius?}

We have seen that our analysis naturally leads us to redefine halo boundaries.  Recently, the so-called splashback radius has been proposed as a physical halo boundary \citep{diemerkravtsov14, More2015}.  We compare the halo radius we derive to the splashback radius as defined using the \sparta\ algorithm \citep{diemer2017, diemeretal2017}.  \sparta\ tracks the orbits of all particles in a halo and measures the location of the first apocenter of all particles.  The splashback radius of a halo is defined as the smoothed average of the apocenter radii of a fraction of the particles.  Common choices are the 75th and 87th percentiles, which roughly match the splashback radius defined as the steepest point of the logarithmic slope of the spherically averaged density profile \citep{More2015}, and as the radius of the sphere with volume equal to the splashback shell of a halo, as first introduced in the code {\tt SHELLFISH} \citep{mansfield2017}, respectively.  When computing the splashback radius of a halo in the simulation, we rely on the $M_{\rm 200m}$ mass of the halo as measured in the simulation.

\begin{figure}
\hspace{-12pt} \includegraphics[width=80mm]{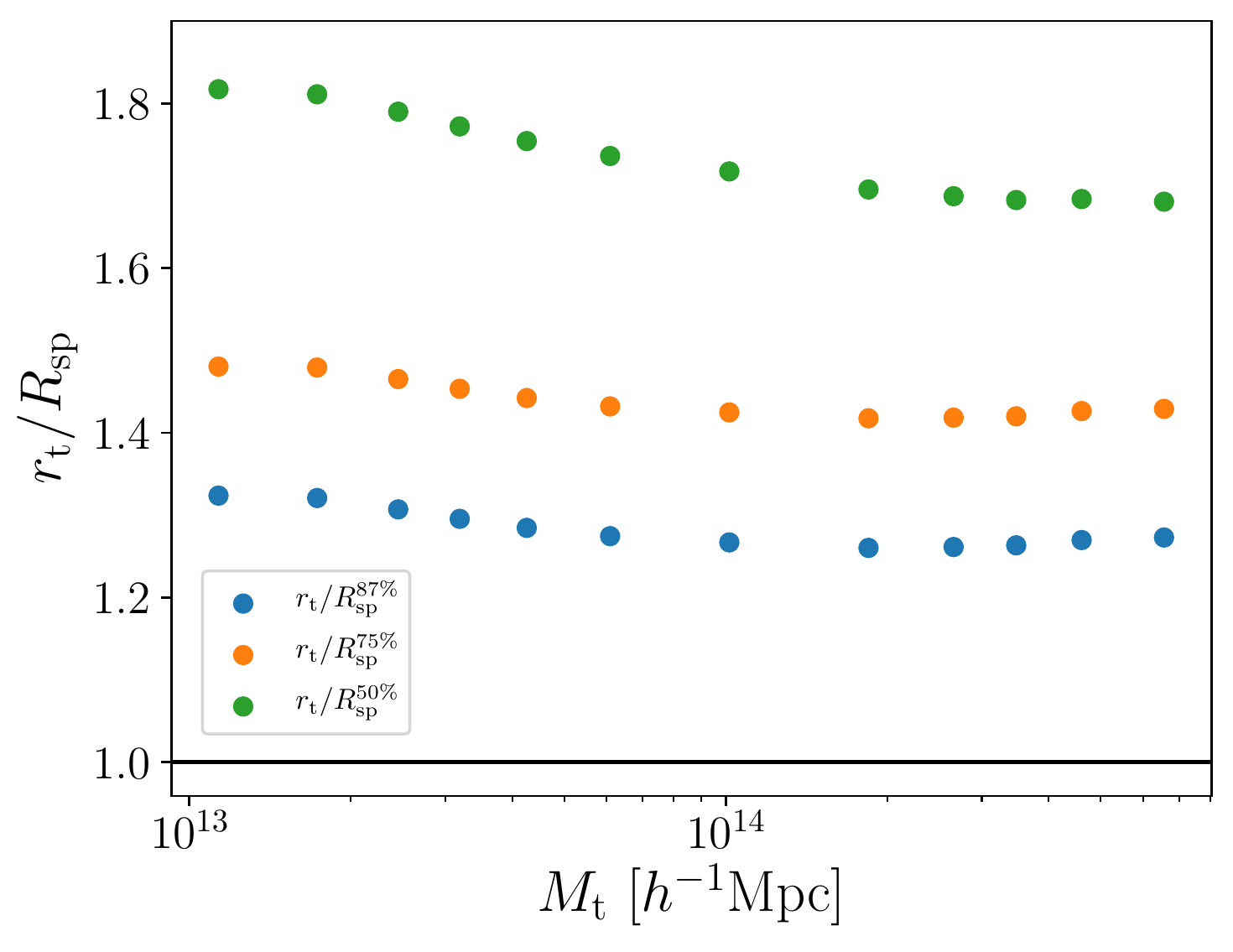}
\caption{Ratio between our proposed halo boundary $\rt$ and the splashback radii from the \sparta\ algorithm for a variety of percentiles of the apocenter distribution of dark matter particles. The ratio seems to be roughly constant.  We believe that \it for the halo population we identified \rm the \sparta\ estimates of the splashback radius becomes increasingly biased as we move to lower masses.  See text for further details.}
\label{fig:rt-rsp}
\end{figure}

Figure \ref{fig:rt-rsp} shows the ratio $\rt/R_{\rm sp}$ for several splashback definitions, specifically the median, 75th and 87th percentiles.  For each mass bin, the splashback radius is the average $R_{\rm sp}$ of all halos in that bin, as estimated from the $M_{200\rm m}$ masses of the halos using the \texttt{SPARTA} code \citep{diemer2017}.  We see that these ratios are roughly constant throughout the mass range $[10^{13}, 10^{15}]\ \hmsun$.
Taking the 87-percentile splashback radius as our reference, we find that $\rt/R_{\rm sp}\approx 1.3$.  This value is close to but somewhat smaller than the edge radius $R_{\rm edge}/R_{\rm sp} \approx 1.55$ identified in \citet{Aung-phasespace}.  It is interesting that both the edge radius and the radius $\rt$ defined here are roughly constant factors of the splashback radius, and that they are both somewhat larger than the splashback radius.  We leave a detailed analysis of how these two different radial scales are related to future work.

As we move to smaller masses, the ratio $\rt/R_{\rm sp}$ grows. We caution, however, the the splashback radii measured at low masses as estimated from \sparta\ are likely biased \it for our halo population, \rm  with the bias almost certainly increasing with decreasing mass.   To see this, recall that \sparta\ was calibrated using parent halos identified with the \rockstar\ halo finder using $R_{\rm 200m}$ as the halo radius.  Since our halo boundary is significantly larger than $R_{\rm 200m}$, a low-mass halo neighboring a high mass halos will become a substructure of the high mass halo upon applying our new halo definition.  These ``halos'' currently contribute to the estimates in \sparta, but are not included in our analysis as parent halos due to the change in percolation in our halo catalog. In other words, the halo population in which \sparta\ was calibrated does not match our halo population, except at the very highest masses.  This implies that a proper comparison of the splashback radius to our proposed halo boundary $\rt$ requires recalibration of the particle orbits based on the halos identified by our algorithm only.  We defer this recalibration to future work.

In short, we believe that splashback radii, the halo edge proposed in \citet{Aung-phasespace}, and the truncation radius we identified as naturally arising from the halo--mass correlation function are all related, though exactly what this relation is remains unclear.  Clarifying the relation between these radii is ongoing work.


\subsection{The Halo Mass Function}

The change in halo definition we suggest directly impacts the halo mass function.  We measured the halo mass function of the final halo catalog produced by our iterative algorithm. The extended Press-Schechter formalism \citep{press-schechter1974} leads to a theoretical prediction of the halo mass function of the form
\begin{align}
\frac{dn}{dm} &= f(\sigma) \frac{\mean{\rho}_{\rm m}}{m} \frac{d \ln \sigma^{-1}}{dm}
\end{align}
where $f(\sigma)$ is some function, and $\sigma(M)$ is the variance of the linear density field over an aperture $\rt(M)$.  Press and Schechter \citep{press-schechter1974} derived a first expression for $f(\sigma)$ on the basis of the spherical collapse model.  The Press-Schechter multiplicity function $f(\sigma)$ is given by 
\begin{align}
f(\sigma) &= \sqrt{\frac{2}{\pi}} \frac{\dsc}{\sigma} \exp \left[ -\frac{\dsc^2}{2\sigma^2} \right]
\end{align}
where $\dsc$ is the critical density required for spherical collapse.  At $z=0$, and assuming a matter density $\Omega_{\rm m}=1$, one finds $\dsc=1.686$.   The quantity $\nu \equiv \dsc/\sigma(M)$ is typically referred to as the peak height.
\\

\begin{figure*}
\hspace{-12pt} \includegraphics[width=80mm]{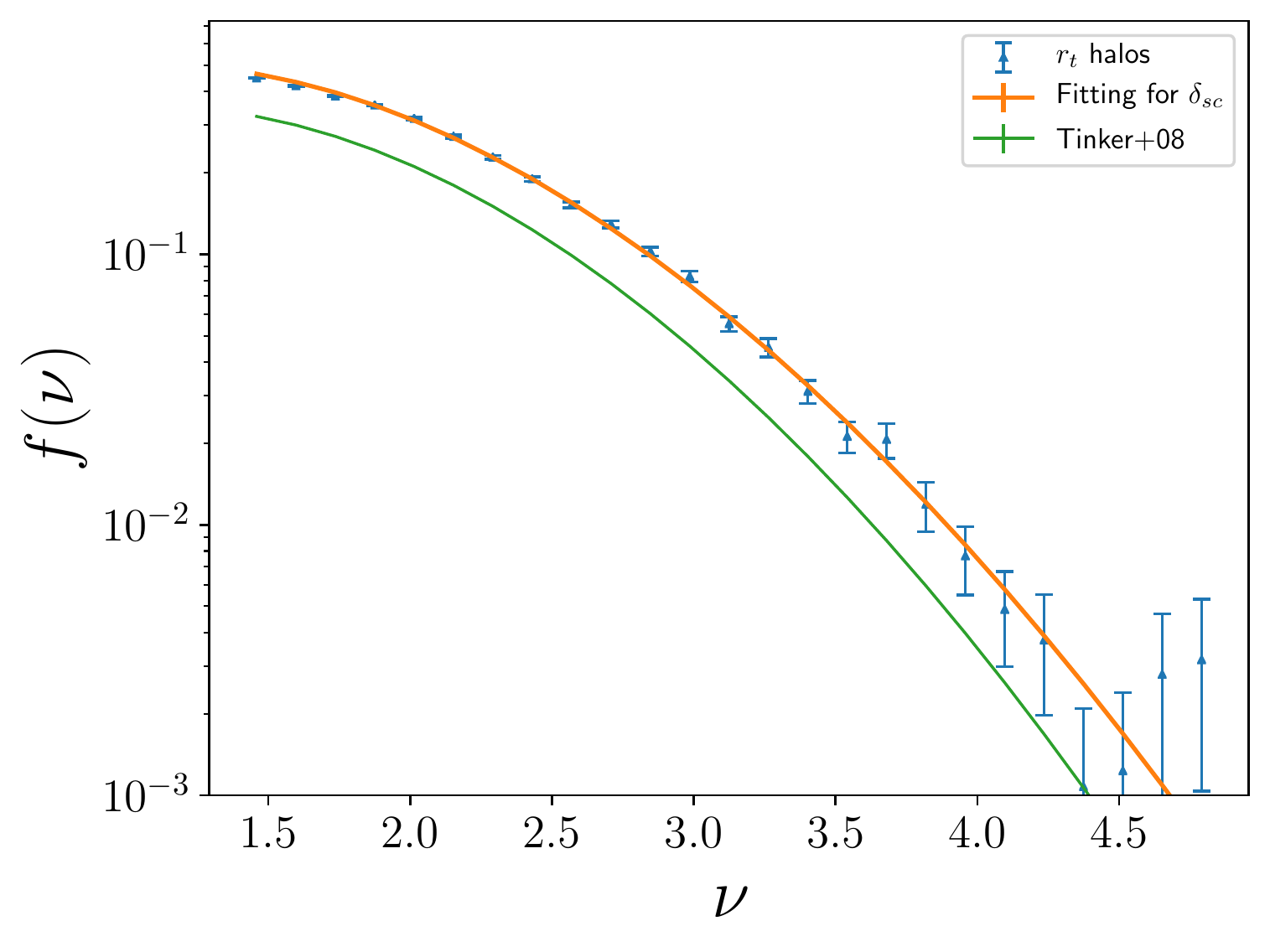}
\includegraphics[width=80mm]{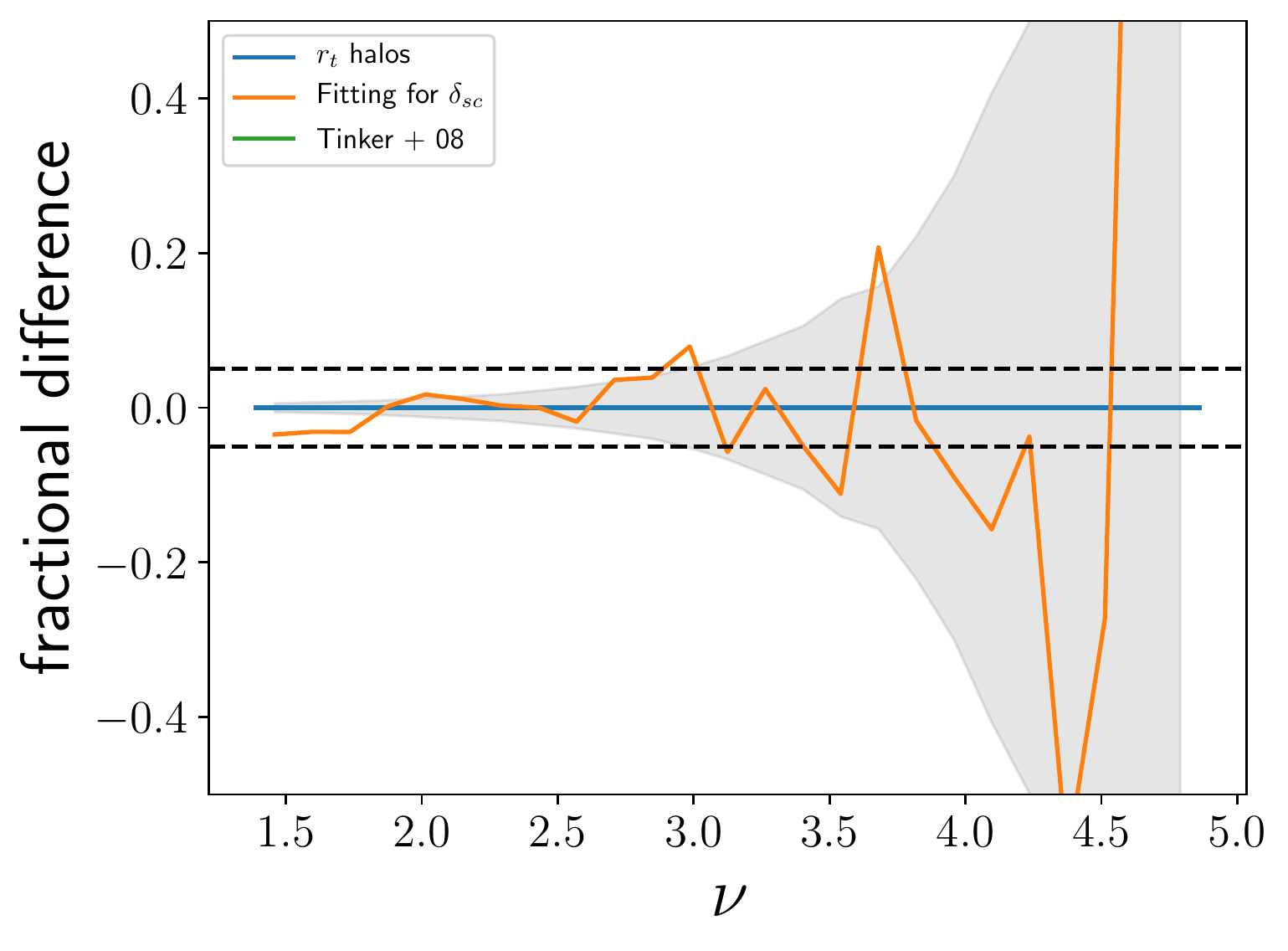}
\caption{{\bf Left panel:} The halo mass function in the simulation using our proposed halo definition (blue points with error bars) and the best fit Press--Schechter mass function (orange line).  The \citet{Tinker2008} mass function (green line) is shown for reference.  Halo mass functions plotted as a function of peak height $\nu$.  Error bars are jackknife.  {\bf Right panel:} Residuals between the simulation data and the best fit Press--Schechter model. The horizontal dashed lines correspond to 5\% deviations.
}
\label{fig:hmf}
\end{figure*}

As shown in Figure~\ref{fig:hmf}, we find that the Press--Schechter mass function gives an excellent fit ($\sim 5\%$ precision) to the mass function of our final halo catalog, provided we fit for the value of $\dsc$.  The posterior on the critical density for collapse $\dsc$ from our best fit Press--Schechter model is $\dsc=1.449 \pm 0.004$.  The excellent agreement between the simulation and the Press--Schechter mass function is surprising, as our analysis did not make any assumptions about halo abundances.  Rather, it relied exclusively on features in the halo--mass correlation function to motivate the redefinition of halo boundaries.  The best fit critical threshold for spherical collapse $\dsc$ is smaller than expected \citep[e.g.][]{paceetal17}.  Whether this specific value can be predicted theoretically remains to be seen.

\subsection{Halo Bias and The Peak--Background Split}

We have shown that the theory of spherical collapse can accurately predict the halo mass function in a simulation, provided we use the correct halo definition and fit for the value of the critical overdensity.  In this section we test whether the peak--background split model of halo bias provides an equally accurate description of our data.  The peak--background split predicts the bias as a function of peak height is given by \citep{mowhite1996, colekaiser1989}
\begin{align}
    \bPB (\nu) = 1 + \frac{\nu^2 - 1}{\dsc}
    \label{eq:halobias}
\end{align}

\begin{figure*}
\includegraphics[width=80mm]{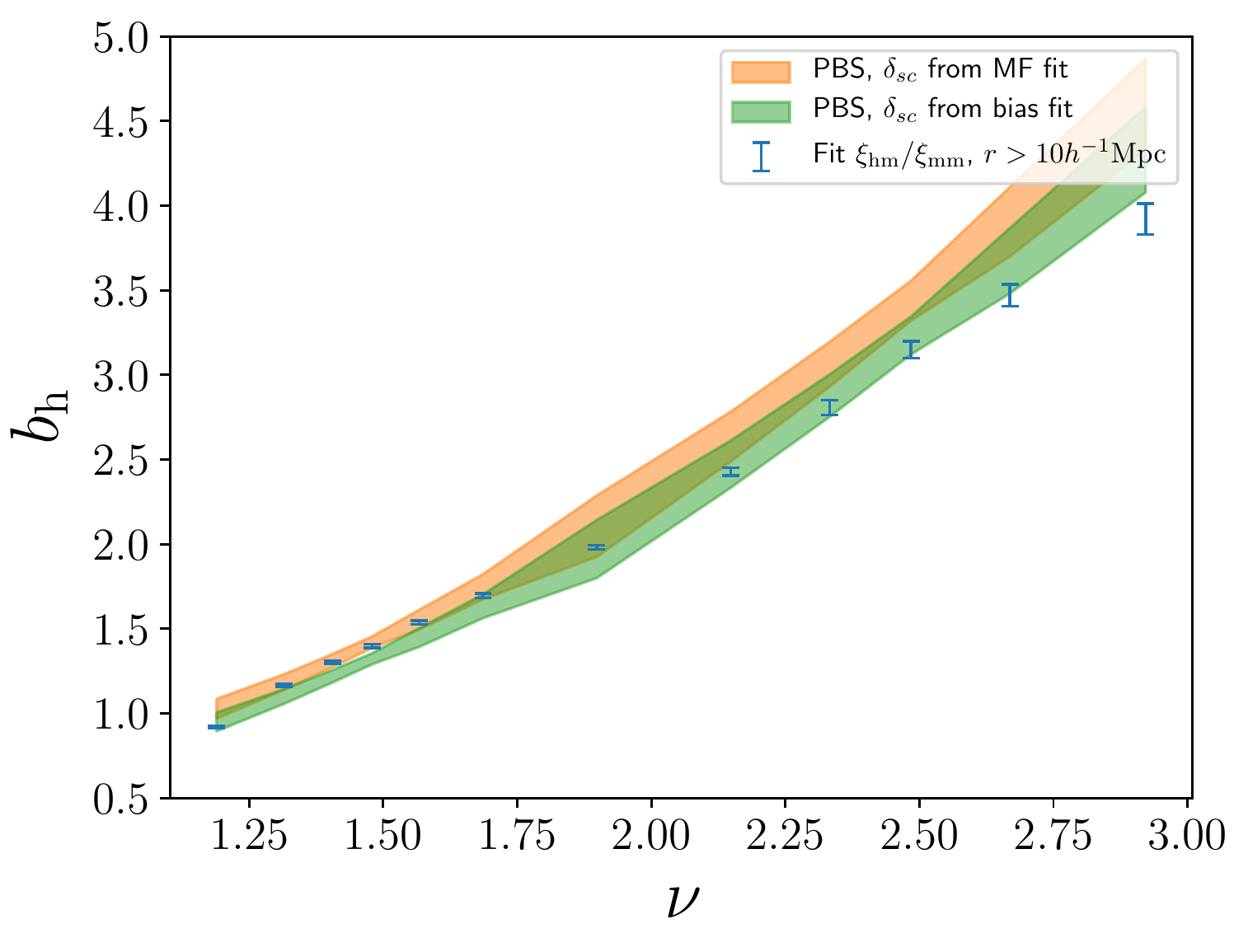}
\includegraphics[width=80mm]{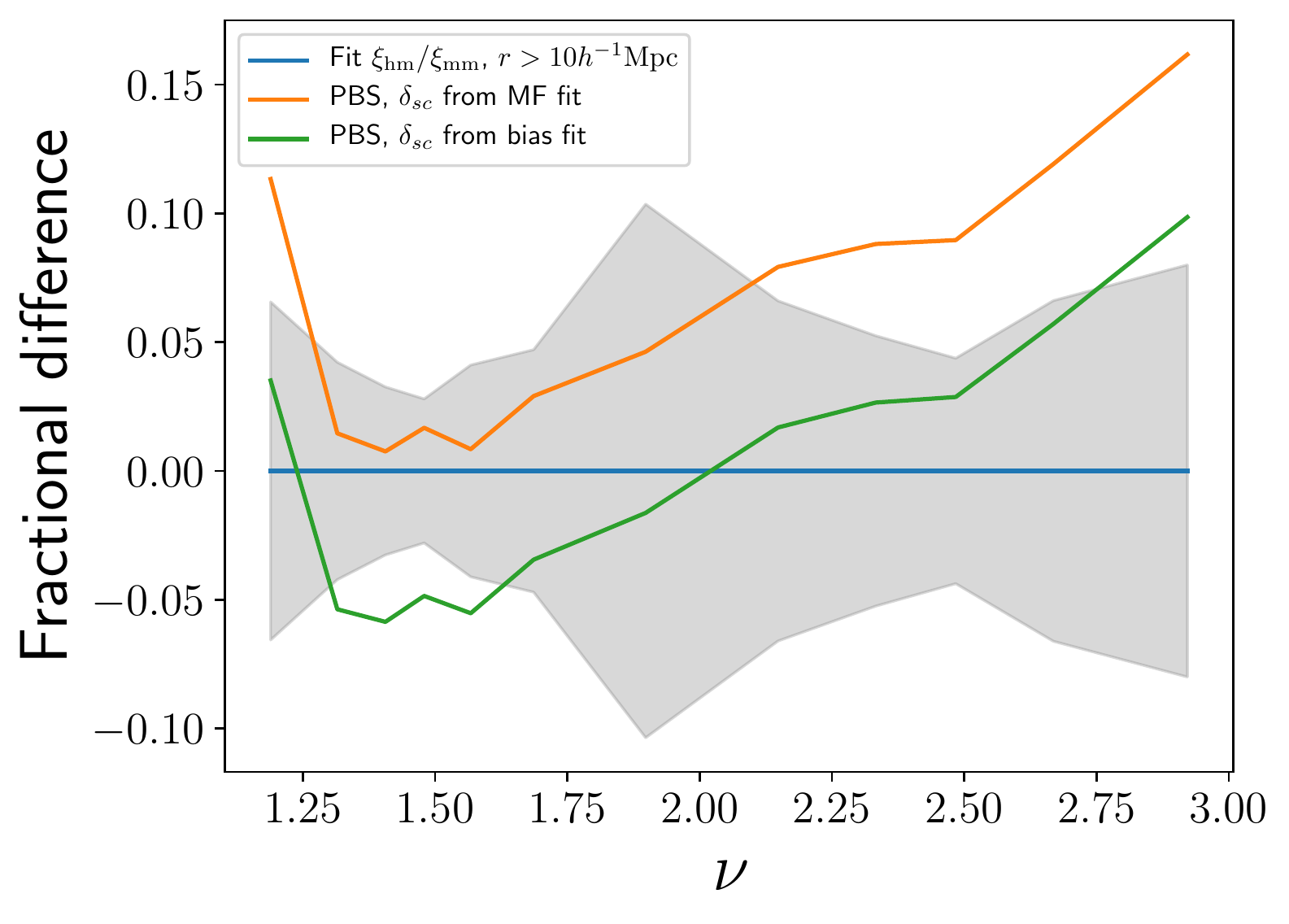}
\caption{The halo bias measured from the ratio of the halo--mass  and mass-mass correlation functions and the prediction from the peak-background split approach.  \textbf{Left:} The halo bias measured from fitting the halo--mass correlation function is shown in blue.  The halo bias calculated using peak-background split prediction with $\delta_{\rm sc}$ the best fit value from the halo mass function fit is shown in orange.  The green curve shows the peak-background split fit where we vary $\delta_{\rm sc}$. \textbf{Right:} Fractional difference with respect to the halo bias $b=\xihm/\ximm$.}
\label{fig:halobias}
\end{figure*}

We calculated the halo bias using the previous equation, where $\nu$ is the peak height as defined in the previous section.  Figure \ref{fig:halobias} shows a comparison between the halo bias as measured using the halo--mass correlation function, and the halo bias derived from the peak background split.  The orange band shows the prediction based on our Press--Schechter fit to the halo mass function.   We see the peak--background split model is roughly $\sim 10\%-15\%)$ consistent with the data, a level of accuracy comparable to the performance of the peak background split for other halo mass definitions \citep[e.g.][]{tinkeretal10,Hoffmann2015,desjacquesetal18}. However, the predicted bias is clearly too high.  We fit our data with a bias of the form derived from the peak--background split, but allow $\dsc$ to vary independently, finding $\dsc = 1.375 \pm 0.012$.  This model can describe our data with $\approx 5\%$ accuracy, though the residual clearly exhibit structure as a function of peak height.  Note than when evaluating the bias model in equation \ref{eq:halobias}, we vary $\delta_{\rm sc}$ both in the denominator and in the definition of the peak height $\nu$.  This is obviously inconsistent with the fit from the halo--mass function, but can be thought of simply as a useful empirical fitting function.

The excursion set peaks  formalism of \citet{paranjapeetal13} has succeeded in presenting a framework capable of jointly fitting the abundance and bias functions of dark matter halos within the context of the peak--background split hypothesis.  Critical to this success is the adoption of a mass dependent stochastic barrier for collapse.  Testing whether or not this formalism can successfully account for the discrepancy between our bias measurements and the peak--background split prediction is beyond the scope of this work.

\section{Summary and Conclusions}

We presented a model for the halo--mass correlation function that assumes a scale independent bias and explicitly incorporates halo edges and halo exclusion. We emphasize that all the qualitative features in our model are well motivated a priori.  The specific parameterizations used to implement these features are arbitrary (e.g. one could replace complementary error functions by Fermi-Dirac functions), but their qualitative form are not.  Importantly, our model contains a single scale-independent bias parameter.  The ratio of $\xihm$ to $\ximm$ does have a scale dependence, but this scale dependence is entirely accounted for by the modifications to the naive halo model due to softly-truncated halo profiles, halo exclusion, and the different one-halo terms of $\xihm$ and $\ximm$.

Our main findings can be summarized as follows:
\begin{itemize}
    \item We derived the model proposed by \cite{HayashiWhite2008} from first principles, and showed that this model is the high mass limit of a halo model that incorporates halo edges and halo exclusion.

    \item Halo exclusion introduces corrections in the halo--mass correlation function at the translinear regime. 
    
    \item In our model, there is a unique radius--mass power-law relation that can be used to define halos for which our model provides an accurate ($\approx 2\%)$ description of the halo--mass correlation function across a wide range of scales.
        
    \item The halo radius identified in our analysis is located at the ``by eye'' transition from the one--halo term to the two--halo term.
    
    \item The halo radius identified in this paper and the splashback radius (calibrated with $R_{200 \rm m}$ halos) are related by a roughly constant multiplicative factor.   However, the exact relation between these two scales, and the edge radius advocated by \citet{Aung-phasespace}, remains unclear, and is the focus of ongoing work.
    
    \item The mass function of halos defined using the halo radius identified in this work is well described by the Press--Schechter formula, though the best fit value for the critical density for spherical collapse $\dsc$ ($\dsc=1.449\pm 0.004$) is below its expected value $\dsc \approx 1.686$.
    
    \item The halo bias prediction from the peak-background split approach are not consistent with the halo bias measured from the simulation, exhibiting 10\% to 15\% offsets depend on halo mass. These differences are comparable to the deviations from the peak--background split prediction for more traditional fixed-overdensity halo definitions.  Remarkably, however, if we independently fit for $\dsc$ in the halo bias expression derived from the peak--background split, we find that a model with $\dsc=1.375\pm 0.012$ can describe our data with $\approx 5\%$ accuracy.
\end{itemize}


It is very encouraging that multiple lines of evidence are now pointing towards the existence of a true halo boundary that extends well beyond $R_{200{\rm m}}$, even if the precise relation between these works is still unclear.  Encouragingly, we have shown that defining halos using our proposed halo boundary significantly simplifies the halo model while improving accuracy.  When coupled with new insights into the halo model, we may soon arrive at a complete theory of large scale structure capable of describing observations at all scales, with the necessary precision required to make full use of upcoming photometric and spectroscopic surveys.  Such an analytic model may appear quaint given the existence of emulators and simulation-rescaling techniques capable of making high accuracy predictions \citep{nishimichietal19,anguloetal20}.  However, we believe there remains significant value to the insights gained from our analytic treatment. To paraphrase Eugene Wigner, it is nice that computers can understand the problem, but we would like to understand it too.

{\it Acknowledgements:} ER and RF were supported by the DOE grant DE-SC0015975.  RG is also supported by CONACyT scholarship 710106.  ER also acknowledges funding from the Cottrell Scholar program of the Research Corporation for Science Advancement. This research was supported by the Munich Institute for Astro- and Particle Physics (MIAPP) which is funded by the Deutsche Forschungsgemeinschaft (DFG, German Research Foundation) under Germany's Excellence Strategy - EXC-2094 - 390783311.  The authors would like to thank Han Aung for useful comments on an early version of this manuscript.  ER would like to thank Ravi Sheth, and Bhuvnesh Jain for useful discussions and suggestions regarding the content of this work. 
  
\bibliographystyle{mnras}
\bibliography{database.bib} 



\label{lastpage}

\end{document}